\documentclass[twocolumn,preprintnumbers,nofootinbib,amsmath,amssymb,floatfix,showpacs]{revtex4}
\usepackage{graphicx}
\usepackage{dcolumn}
\usepackage{bm}

\def\neff{N_{\rm eff}}
\def\he4{$^4$He}
\def\h2{$^2$H}
\begin{document}

\title{A robust upper limit on $N_{\rm eff}$ from BBN, circa 2011}

\author{Gianpiero Mangano}\email{mangano@na.infn.it}
\affiliation{LAPTh, UMR 5108, 9 chemin de Bellevue - BP 110, 74941 Annecy-Le-Vieux, France}
\affiliation{INFN, Sezione di Napoli, Complesso Univ. Monte S. Angelo, Via Cintia, I-80126 Napoli, Italy}
\author{Pasquale D.~Serpico}\email{serpico@lapp.in2p3.fr}
\affiliation{LAPTh, UMR 5108, 9 chemin de Bellevue - BP 110, 74941 Annecy-Le-Vieux, France}

\begin{abstract}
We derive here a robust bound on the effective number of neutrinos from constraints on primordial nucleosynthesis yields of deuterium and helium. In particular,
our results are based on very weak assumptions on the astrophysical determination of the helium abundance, namely that the minimum
effect of stellar processing is to keep constant (rather than increase, as expected) the helium content of a low-metallicity gas. Using the results of a recent analysis of 
extragalactic  HII regions as {\it upper limit}, we find that $\Delta\neff\leq 1$ at 95\% C.L., quite independently of measurements on the baryon density from cosmic microwave background anisotropy data and of the neutron lifetime input. In our approach, we also find that  primordial nucleosynthesis alone has no significant preference for an  effective number of neutrinos larger than the standard value. The $\sim 2\,\sigma$ hint sometimes reported in the literature is thus driven by CMB data alone and/or is the result of a questionable regression protocol to infer a {\it measurement} of primordial helium abundance.
\end{abstract}

\pacs{26.35.+c
\hfill LAPTH-004/11}
\maketitle

\section{Introduction}

Historically, the helium abundance has played an important role for establishing the ``hot big bang'' cosmological model, see for example~\cite{Kra96}. Primordial 
nucleosynthesis, most often referred to as big bang nucleosynthesis (BBN), helped building confidence in the overall credibility of cosmology as a science. Nowadays, in the
era of ``precision cosmology'', it is clearly no more question to prove that the bulk of \he4 is primordial, also in view of the indirect (but clean) detection of a non-vanishing $^4$He mass fraction $Y_p$ from CMB data, see~\cite{Komatsu:2010fb,Dunkley:2010ge}.

The basic pillars of the hot big bang model have found many confirmations and have been subject to several important cross-checks, leading to the so-called concordance (or standard) model of cosmology. In the last two decades,  BBN has thus mostly turned into a probe of physics/cosmology beyond the standard model (see~\cite{Iocco:2008va,Pospelov:2010hj} for recent reviews). Moving forward in this direction has been however hampered by the systematics in the determination of light nuclei abundances, since one has no access to truly primordial environments, i.e. prior to the first generation of stars and nucleosynthetic events: any inference thus relies to some extent on  astrophysical models. In particular for the case of \he4, despite the more and more careful analyses undertaken over the past decade, very different conclusions can be sometimes found  on the inferred  $Y_p$, see e.g.~\cite{Izotov:2010ca,Aver:2010wq,Aver:2010wd}. Without entering the issues related to the analysis of spectroscopic data, from a particle physics perspective one may wonder what is the most profitable way to use the results of these analyses for constraining new physics, while maintaining some robustness and independence from the regression protocol. 

In fact, we believe that for a largest fraction of the particle astrophysics community it is more important to obtain reliable constraints than inferring primordial abundances. In this spirit, here we propose a simple and robust approach to the use of \he4 data for constraining the  effective number of neutrinos, $\neff$, by far the most widely used BBN-related quantity used  to parameterize new physics. This Letter is structured as follows: in Sec.~\ref{procedure} we outline our minimal assumptions in using the data; in Sec.~\ref{results} we present our results; in Sec.~\ref{discconcl} we discuss our findings and assumptions and conclude.

\section{Procedure}\label{procedure}
For \he4,  we use here the abundances inferred in nine metal-poor, extragalactic HII regions in~\cite{Aver:2010wd} (see also~\cite{Aver:2010wq}).
Differently from the usual practice, we do not perform a regression to zero metallicity, since our aim is to derive an upper bound to \he4. Hence,
we fit the yields to a constant abundance $Y_0$, obtaining 
\begin{equation}
\langle Y_0\rangle\pm \sigma_0=0.2581\pm 0.0025\, (68\%\:{\rm C.L.})\,.\label{maxfit}
\end{equation}
which also implies
\begin{equation}
Y_0< 0.2631\:\: {\rm at}\:95\%\:{\rm C.L.}\,\label{upper}
\end{equation}
Assuming that no information is available on the lower-limit of \he4, one can parametrize the likelihood function for $Y_p$ as a flat+semi-gaussian shape
\begin{equation}
\ell(Y_p)\propto \Theta(\langle Y_0\rangle-Y_p)+\Theta(Y_p-\langle Y_0\rangle)\exp\left[-\frac{(Y_p-\langle Y_0\rangle)^2}{2\,\sigma_0^2}\right]\,.\label{likeHe4}
\end{equation}
 
 We argue here that this is conservative and robust. In fact, 
Eq.~(\ref{upper}) is a strict upper limit to the primordial value $Y_p$ under {\it the sole assumption} that $dY/dZ\geq 0$, i.e. that the {\it average}~\footnote{Note that  short periods of evolution with $dY/dZ< 0$ would not alter the conclusion, hence the requirement is less restrictive that the inequality $dY/dZ\geq 0$ taken at face value.}
effect of stellar processing is to increase the helium content of a low-metallicity gas. By now, there is some empirical
evidence of this trend (positive derivative) even in the few high-quality objects analyzed in~\cite{Aver:2010wd,Aver:2010wq}.
However, assuming $dY/dZ=$constant is a much stronger assumption, usually justified empirically since the seventies as the simplest possible fit~\cite{P74} (see also~\cite{Fields:1994nc}), but subject to the obvious risk of extrapolation errors.

Even if one could model the metallicity evolution of the observed systems reliably, it is unclear to what extent the {\it pre-galactic} value of $Y$ should coincide with its {\it primordial} value $Y_p$.  Actually, it has been proven since longtime that the bulk of \he4 must
be primordial (see e.g.~\cite{Peimbert:2008bt} for a short historical overview), and by now there is a positive (albeit indirect) detection of a non-vanishing $Y_p$ from
CMB data, see~\cite{Komatsu:2010fb,Dunkley:2010ge}. However,  it is a different issue to prove that no extra production of  \he4 has taken place since early times  {\it at a level comparable to the current  error} on its determination, i.e. at the percent level. In fact, 
``What is the helium enrichment  by the first stellar generations?'' was a question already posed in the seminal
paper~\cite{hoyle64}. In the last decade, several models have been proposed where such a production does take place
at an appreciable level ($\Delta Y\simeq 10^{-3}-10^{-2}$), see for example~\cite{Salvaterra:2003nu,Vangioni:2010wf}. This is perhaps not surprising, given that the yet undiscovered generation of stars known as PopIII forming in the pristine metal-free gas should generate some chemical enrichment. It is worth remembering that currently observed ``metal-poor'' samples have metal contents many orders of magnitude above the expectations from  BBN~\cite{Iocco:2007km}. When using the 
late universe determinations of $Y$ as measurements of its primordial value, a systematics error
is committed, difficult to quantify precisely since it depends on astrophysical modeling of pre-galactic times. 
Again, this extra problem is avoided in our conservative procedure.
Notice that our upper bound of Eq.~(\ref{upper}) is close but slightly more stringent than what is found at 2 $\sigma$ in~\cite{Aver:2010wd}, $Y_p < 0.2639$, since only seven of the nine determinations were used there.  The small difference shows however that a minor change in the choice of objects would not change the following results appreciably, which is another hint in favor of their robustness. It is more involved to compare with the results of the group of Izotov et al., which does not present fits
assuming $dY/dZ= 0$. For illustration, the {\it measurement} quoted in~\cite{Izotov:2010ca}, based on 
a linear regression method whose robustness we question here, yielded $Y_p=0.2565\pm 0.0010\,(1\sigma\,\,{\rm stat.})\pm 0.0050\,$(syst.). On the other hand,  we note that fitting to a constant the 10 determinations reported in Table 4 of~\cite{Izotov:2007ed}, as suggested here, one obtains $Y_0=0.2555\pm0.0016$,
i.e. $Y<0.2587$ at 95\% C.L.  When  rescaling the above value upwards by 2\% due to some improved atomic
corrections as argued in~\cite{Izotov:2010ca}, one finds a value remarkably consistent with the above determinations of the upper limit.
We take this exercise as an indication that our results are not sensitive to the different analysis codes and protocols, at least when the same and most updated atomic data are used.
 
For \h2, following the analysis in~\cite{Iocco:2008va}, as best estimate of the primordial deuterium yield one may use
\begin{equation}
({}^2{\rm H}/{\rm H})_p=2.87_{-0.21}^{+0.22}\:\: {\rm at}\:68\%\:{\rm C.L.}\,\label{deut}
\end{equation}
which is based on a conservative analysis of seven (relatively) reliable absorption spectra of clouds at high redshifts, on the light of background quasars. 
Detailed studies~\cite{Fields:1994nc} suggest that the depletion due to stellar activity in these early systems is negligible and thus the
above range can be considered a faithful estimate of the primordial value. Yet, we shall also show one example based on the minimal
assumption 
\begin{equation}
({}^2{\rm H}/{\rm H})_p>2.45\times 10^{-5}\:\: {\rm at}\:95\%\:{\rm C.L.}\,,\label{deutLOW}
\end{equation}
which is agnostic on an eventual depletion of ${}^2{\rm H}$ in an early chemical evolutionary phase. In this case, a semigaussian+flat likelihood function  similar to what done for Helium in Eq.~(\ref{likeHe4}) is constructed. Even in this case, very similar results follow on the upper limit on $\neff$.

Concerning the BBN predictions, we used nuclide yields based on the code described in~\cite{Pisanti:2007hk}, also including the uncertainties due to nuclear reactions extensively described in~\cite{Serpico:2004gx}.

Finally, one may (or may not) impose the CMB measurement of the baryon fraction~\cite{Komatsu:2010fb}
\begin{equation}
\omega_b=\Omega_b\,h^2=0.02250\pm0.00056\:\: {\rm at}\:68\%\:{\rm C.L.}\,\label{WMAP}
\end{equation}
We checked that significantly relaxing the above range (say, by one order of magnitude) does not affect the upper limit on $\neff$ appreciably. Note that for consistency we consider the value for $\omega_b$ inferred in a $\Lambda$CDM+$\neff$ model (i.e. where $\neff$ is allowed to vary), although using the best-fit in a $\Lambda$CDM model would make no quantitative difference.

For illustration, in one case we shall also consider the impact of the $Y_p$ measurement from CMB data (WMAP 7 yrs plus small scale results of the
Atacama Cosmology Telescope~\cite{Dunkley:2010ge}), assuming the additional measurement (gaussian error)
\begin{equation}
Y_p=0.313\pm0.044 \:\: {\rm at}\:68\%\:{\rm C.L.}\,\label{Ycmb}
\end{equation}

\section{Results}\label{results}
With the ingredients described above we constructed two-dimensional likelihood functions  in the $\omega_b-\neff$ plane, then marginalized
over the parameter $\omega_b$, which is not of interest here. The results of our analysis are thus encoded in the 1-dimensional likelihoood functions ${\cal L}(\neff)$, whose integrals are normalized to 1. These functions are shown in Fig.~\ref{likeli} and relevant numerical quantities are summarized in Table~\ref{tab1}. We define $\neff^{\rm min}$ and $\neff^{\rm max}$ such that
\begin{equation}
\int_{\neff^{\rm min}}^7 {\cal L}(x)d\,x=0.95\,\,\:,\:\:\:
\int_0^{\neff^{\rm max}} {\cal L}(x)d\,x=0.95\,,
\end{equation}
and the parameter $L(\neff \leq\neff^{\rm SM})$ in Table~\ref{tab1} as
\begin{equation}
L(\neff \leq\neff^{\rm SM}) =\int_{0}^{\neff^{\rm SM}} {\cal L}(x)d\,x\,.
\end{equation}

{\begin{table}[t]
\begin{tabular}{||l||r|r|r||}\hline
Datasets & $\neff^{\rm max}$ & $\neff^{\rm min}$ & $L$($\neff \leq\neff^{\rm SM}$)   \\
\hline\hline 
$\omega_b$+\h2+\he4 & 4.05 & 2.56 & 0.20  \\
\hline
$\omega_b$+\h2$_{\rm low}$ +\he4 & 4.08 & 2.57 & 0.19  \\
\hline
\h2+\he4 & 3.91 & 0.80 & 0.67  \\
\hline
$\omega_b+Y_p^{\rm CMB}$+\h2+\he4 & 4.08 & 2.71 & 0.15  \\
\hline
\end{tabular}
\caption{Constraints on $\neff$ corresponding to different datasets used: i) first row: Eq.~(\ref{WMAP}), Eqs.~(\ref{maxfit};\ref{likeHe4}), Eq.~(\ref{deut}); ii) second row: Eq.~(\ref{WMAP}), Eqs.~(\ref{maxfit};\ref{likeHe4}), Eq.~(\ref{deutLOW});
iii) third row: Eqs.~(\ref{maxfit};\ref{likeHe4}) and Eq.~(\ref{deut}); fourth row: as the first one, with the additional CMB measurement of $Y_p$ of Eq.~(\ref{Ycmb}). The last column shows the likelihood that $\neff$ is smaller than the standard value 3.046~\cite{Mangano:2005cc}. }\label{tab1}
\end{table}
When remembering that the  standard model expectation for $\neff$ is about 3.046~\cite{Mangano:2005cc}, we see that in all cases we get a bound  $\Delta \neff\leq 1$. The reason why it is slightly more stringent when using deuterium as a ``measurement'' of $\omega_b$
instead of CMB (third line) is that it favors a slightly smaller value for the baryon fraction. In correspondence of this smaller value, the deuterium yield is a bit larger. Since deuterium
{\it grows} with $\neff$, in this case the deuterium hits the upper bound for a lower value of $\neff$, increasing its constraining power.
As it is clear from the second row of Table~\ref{tab1}, allowing for primordial deuterium depletion and limiting oneself to consider the lowest limit of its measured value as a lower
limit, the bound does not change much, since the constraining power derives from the upper limit on \he4. In Fig.~\ref{likeli}, this reflects on the quite hard cut in the likelihood functions
at large $\neff$. Also, adding the CMB measurement of $Y_p$ of Eq.~(\ref{Ycmb}) does not change much the situation with respect to the first case: the slight shift towards higher values of $\neff$ reported in the fourth row is simply due to the fact that the current best value of $Y_p$ from CMB is above the BBN prediction, albeit not significantly (less than 1.5 $\sigma$). This also proves indirectly that if we had imposed a loose lower-bound  on $Y_p$ (say, $Y_p>0.225$) instead of the flat likelihood of Eq.~(\ref{likeHe4}) at low-$Y_p$,
the result would hardly change.

On the other hand, comparing the first and last two  lines in the table shows that an independent constraint on $\omega_b$ and possibly even a relatively weak lower limit on $Y_p$ are quite useful in setting a stringent {\it lower} limit on $\neff$ (second column of Table~\ref{tab1}). In particular, the effect of the constraint on $\omega_b$ is explained as follows:  since the dependence of  \he4 on $\omega_b$ is very weak, and \h2 suffers of a partial degeneracy between $\neff$ and $\omega_b$, relatively low values of $\neff$ can be compensated with
relatively high values of $\omega_b$. Hence, imposing an upper limit on $\omega_b$ yields to a more stringent lower limit on $\neff$. Of course,  this exercise has only illustrative purpose: the physics behind the CMB measurement on $\omega_b$ is  well understood, and any cosmologically meaningful lower limit on $\neff$ is significantly larger than the value reported at the third row in Table~\ref{tab1}. 
\begin{figure}
\begin{center}
\includegraphics[angle=0,width=0.5\textwidth]{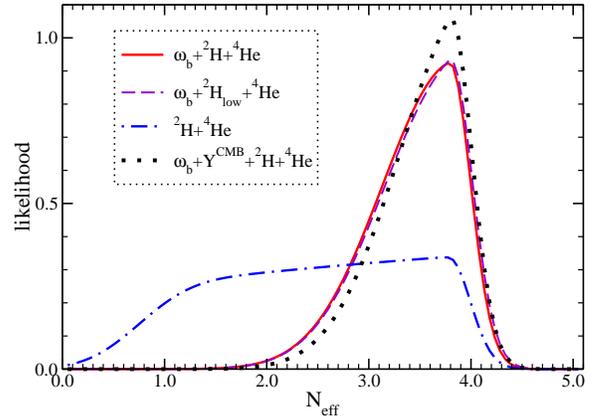} 
    \end{center}
\caption{\label{likeli} Marginalized 1-D likelihood functions $\cal L$  versus $\neff$ using the different combinations of data as in Table \ref{tab1}. Solid (red) and dashed (purple) curves are obtained using CMB measurement of $\omega_b$, with the dotted (black) one also adds CMB information on $Y_p$. In all cases the quite sharp cut-off at $\neff \sim 4$ is due to $^4$He abundance upper limit.}
\end{figure}

Finally, it is interesting to consider the last column in Table~\ref{tab1}, illustrating the likelihood that the inferred $\neff$ value is lower or equal than its standard model expectation: we see that BBN alone  has no clear preference for a larger-than-standard $\neff$ (compared to a lower-than-standard one) when the observed abundances are interpreted conservatively. The blue, dot-dashed curve in Fig.~\ref{likeli} also shows graphically the same effect. Even when combined with CMB data, BBN does not favor significantly larger-than-standard values for $\neff$.

\section{Discussion and Conclusions}\label{discconcl}
In this letter, we have discussed a new and more conservative approach to derive  BBN constraints on $\neff$,  motivated by growing concerns on the reliability of astrophysical determinations of primordial \he4. We showed that even with present data, an analysis which is at the same time conservative and informative is possible,
in particular when using  CMB and/or deuterium measurements as constraints on the baryon abundance. In all the cases we considered, we found that $\Delta\neff\leq 1$ at 95\% C.L.,
and possibly the bounds are slightly more stringent, see Table \ref{tab1}. It is perhaps useful to note that, following hints from neutrino laboratory experiments, models with two sterile neutrinos with relatively large mixings have been recently (re)considered in a cosmological context, see e.g.~\cite{Hamann:2010bk,Giusarma:2011ex}. Although CMB data alone might accommodate or even marginally favor such amount of extra radiation, our results suggest that they are strongly disfavored or excluded by BBN, at least if the two sterile neutrino distributions are close enough to thermal ones. Of course, this is only an example: non-standard values of $\neff$ (and in particular low ones) might even indicate new physics completely unrelated to neutrinos (more details can be found e.g. in~\cite{Iocco:2008va}). 

Let us briefly discuss the sensitivity of our results to some underlying assumptions. Let's assume that our sole astrophysical assumption on helium, $dY/dZ\geq 0$, should be {\it empirically} proven wrong. 
In such a case (at the moment of academic interest, given the evidence of the contrary from the data!), lacking a model-independent way to correct for the evolutionary effects,   the whole logic of ``primordial helium traceability'' would be put under discussion: otherwise said, we expect that observational progresses will either improve over our upper bound (for example by limiting the analysis to very low metallicity objects and/or increasing the statistics), or one will have to give up completely the idea that any quantitative inference on the primordial value can be done from what observed in low-redshift stellar objects. Only indirect probes as the CMB one could then be used reliably, though with quite a large uncertainty.

On the other hand, albeit only in model-dependent frameworks, one has an idea of the possible contributions of pre-galactic generations of stars to $Y$. For example, the model in~\cite{Salvaterra:2003nu} suggest possible production yields of PopIII stars up to $\Delta Y\simeq 3.3\times 10^{-3}$, with comparable effects found in \cite{Vangioni:2010wf}.
From a numerical evaluation of the derivative of $Y_p(\omega_b^{\rm CMB},\neff)$ at the standard value we find $\Delta \neff\simeq 75\, \Delta Y$, hence a contribution like the one above would result in an apparent rightward shift of $\Delta \neff\simeq 0.25$ in the curves of Fig. 1. In~\cite{Vangioni:2010wf} the overall chemical evolution is found to be responsible for shifts up to $\Delta Y\simeq 0.007$ (i.e. $\Delta \neff \gtrsim 0.5$) while the impact on Deuterium abundances in high-redshift objects is negligible. It appears reasonable to conclude that the cosmological evolution $Y(z)$ due to both pre-galactic and galactic populations of objects might easily explain the slight preference for larger-than-standard $\neff$ values from the BBN analysis based on current astrophysical samples. 

One may also wonder how sensitive this bound is to other particle physics parameters entering the \he4 prediction. The most important such parameter is probably the neutron lifetime, which sets the timescale of the neutron/proton density ratio  freeze-out. The two recent determinations~\cite{Serebrov:2004zf,Pichlmaier:2010zz} of the neutron lifetime differ for a  5.6 and 2.5 $\sigma$'s respectively, from the PDG recommended world average~\cite{Nakamura:2010zzi} used in deriving our results above. Yet, we find that adopting the new determinations of the neutron lifetime would only relax the bound by  $\Delta \neff\lesssim $0.1.

As already noted in the past, the synergy of BBN and CMB is useful in narrowing some parameter uncertainties, once again even under very weak/conservative assumptions on the measured helium mass fraction. Forthcoming data from the Planck experiment are expected to strongly increase the sensitivity to $\Delta \neff$, while being insensitive to astrophysical pollution of primordial $Y_p$. Values for this parameter compatible with BBN data expectations, as for example those presented in this letter, could then be further refined by exploiting the theoretical known dependence of the $^4$He mass fraction $Y_p(\omega_b,\neff)$ in the analysis, thus reducing the impact of a flat prior on $Y_p$ on cosmological parameter estimation, see e.g.~\cite{Ichikawa:2006dt,Hamann:2007sb}. On the other hand, evidence for larger than standard  $\neff$ from  CMB would be a real breakthrough, strongly supporting non-minimal  scenarios, as the presence of {\it large} neutrino chemical potentials~\cite{Mangano:2010ei}, or---in case of very large values of $\neff$---hinting for extra species changing the evolution of the Universe possibly after BBN, but before/during CMB formation, one recent example of which has been provided in~\cite{Fischler:2010xz}.
\acknowledgments
G.M. acknowledges support from {\it Universit\'e de Savoie} via its visitor program. We would like to thank A. Melchiorri and G. Raffelt for comments.


\begin{thebibliography}{99}


\bibitem{Kra96} H. Kragh, {\it Cosmology and Controversy}, (1996) Princeton Univ. Press.

\bibitem{Komatsu:2010fb}
  E.~Komatsu {\it et al.}  [WMAP Collaboration],
  ``Seven-Year Wilkinson Microwave Anisotropy Probe (WMAP) Observations:
  Cosmological Interpretation,''
  Astrophys.\ J.\ Suppl.\  {\bf 192}, 18 (2011)
  [arXiv:1001.4538].

\bibitem{Dunkley:2010ge}
  J.~Dunkley {\it et al.},
 ``The Atacama Cosmology Telescope: Cosmological Parameters from the 2008
  Power Spectra,''
  arXiv:1009.0866.

\bibitem{Iocco:2008va}
  F.~Iocco {\it et al.}, 
  ``Primordial Nucleosynthesis: from precision cosmology to fundamental
  physics,''
  Phys.\ Rept.\  {\bf 472}, 1 (2009)
  [arXiv:0809.0631].
  
  
\bibitem{Pospelov:2010hj}
  M.~Pospelov and J.~Pradler,
  ``Big Bang Nucleosynthesis as a Probe of New Physics,''
  Ann.\ Rev.\ Nucl.\ Part.\ Sci.\  {\bf 60}, 539 (2010)
  [arXiv:1011.1054].

\bibitem{Izotov:2010ca}
  Y.~I.~Izotov, T.~X.~Thuan,
  ``The primordial abundance of 4He: evidence for non-standard big bang nucleosynthesis,''
  Astrophys.\ J.\  {\bf 710}, L67-L71 (2010)
  [arXiv:1001.4440].  


\bibitem{Aver:2010wq}
  E.~Aver, K.~A.~Olive and E.~D.~Skillman,
  ``A New Approach to Systematic Uncertainties and Self-Consistency in Helium
  Abundance Determinations,''
  JCAP {\bf 1005}, 003 (2010)
  [arXiv:1001.5218].

\bibitem{Aver:2010wd}
  E.~Aver, K.~A.~Olive and E.~D.~Skillman,
  ``Mapping systematic errors in helium abundance determinations using Markov
  Chain Monte Carlo,''
  arXiv:1012.2385.

\bibitem{P74}
M. Peimbert and S. Torres-Peimbert,
``Chemical composition of H II regions in the Large Magellanic Cloud and its cosmological implications'',
Astrophys.\ J.\ {\bf  193}, 327 (1974).

\bibitem{Fields:1994nc}
  B.~D.~Fields,
  ``On The Evolution Of The Light Elements. 1. D, He-3, And He-4,''
  Astrophys.\ J.\  {\bf 456}, 478 (1996)
  [astro-ph/9512044].

\bibitem{Peimbert:2008bt}
  M.~Peimbert,
  ``The Primordial Helium Abundance,''
  Curr.\ Sci.\  {\bf 95}, 1165 (2008)
  [arXiv:0811.2980].
  
\bibitem{hoyle64}
F. Hoyle and  J. Tayler,
``The Mystery of the Cosmic Helium Abundance''
Nature {\bf 203}, 1108-1110 (1964). 

\bibitem{Salvaterra:2003nu}
  R.~Salvaterra and A.~Ferrara,
  ``Is Primordial He Truly from Big Bang ?,''
  Mon.\ Not.\ Roy.\ Astron.\ Soc.\  {\bf 340}, L17 (2003)
  [astro-ph/0302285].


\bibitem{Vangioni:2010wf}
  E.~Vangioni, J.~Silk, K.~A.~Olive and B.~D.~Fields,
  ``Cosmic Chemical Evolution with an Early Population of Intermediate Mass
  Stars,''
  arXiv:1010.5726.
  
\bibitem{Iocco:2007km}
  F.~Iocco {\it et al.}, 
  ``The path to metallicity: Synthesis of CNO elements in standard BBN,''
  Phys.\ Rev.\  D {\bf 75}, 087304 (2007)
  [astro-ph/0702090].
  
\bibitem{Izotov:2007ed}
  Y.~I.~Izotov, T.~X.~Thuan and G.~Stasinska,
  ``The primordial abundance of 4He: a self-consistent empirical analysis of
  systematic effects in a large sample of low-metallicity HII regions,''
  Astrophys.\ J.\  {\bf 662}, 15 (2007)
  [astro-ph/0702072].
 
\bibitem{Pisanti:2007hk}
  O.~Pisanti {\it et al.},
  ``PArthENoPE: Public Algorithm Evaluating the Nucleosynthesis of Primordial
  Elements,''
  Comput.\ Phys.\ Commun.\  {\bf 178}, 956 (2008)
  [arXiv:0705.0290 [astro-ph]].

\bibitem{Serpico:2004gx}
  P.~D.~Serpico {\it et al.}, 
  ``Nuclear Reaction Network for Primordial Nucleosynthesis: a detailed
  analysis of rates, uncertainties and light nuclei yields,''
  JCAP {\bf 0412}, 010 (2004)
  [astro-ph/0408076].

\bibitem{Mangano:2005cc}
  G.~Mangano {\it et al.}, 
  ``Relic neutrino decoupling including flavour oscillations,''
  Nucl.\ Phys.\  B {\bf 729}, 221 (2005)
  [hep-ph/0506164].


\bibitem{Hamann:2010bk}
  J.~Hamann {\it et al.}, 
  ``Cosmology Favoring Extra Radiation and Sub-eV Mass Sterile Neutrinos as an Option,''
  Phys.\ Rev.\ Lett.\  {\bf 105}, 181301 (2010)
  [arXiv:1006.5276].

\bibitem{Giusarma:2011ex}
  E.~Giusarma{\it et al.}, 
  ``Constraints on massive sterile neutrino species from current and future
  cosmological data,''
  arXiv:1102.4774.

  
\bibitem{Serebrov:2004zf}
  A.~Serebrov {\it et al.},
  ``Measurement of the neutron lifetime using a gravitational trap and a
  low-temperature Fomblin coating,''
  Phys.\ Lett.\  B {\bf 605}, 72 (2005)
  [nucl-ex/0408009].

\bibitem{Pichlmaier:2010zz}
  A.~Pichlmaier, V.~Varlamov, K.~Schreckenbach and P.~Geltenbort,
  ``Neutron lifetime measurement with the UCN trap-in-trap MAMBO II,''
  Phys.\ Lett.\  B {\bf 693}, 221 (2010).
  
\bibitem{Nakamura:2010zzi}
  K.~Nakamura {\it et al.}  [Particle Data Group],
  ``Review of particle physics,''
  J.\ Phys.\ G {\bf 37}, 075021 (2010).


\bibitem{Ichikawa:2006dt}
  K.~Ichikawa, T.~Takahashi,
  ``Revisiting the constraint on the helium abundance from cmb,''
  Phys.\ Rev.\  {\bf D73}, 063528 (2006)
  [astro-ph/0601099].

\bibitem{Hamann:2007sb}
  J.~Hamann, J.~Lesgourgues and G.~Mangano,
  ``Using BBN in cosmological parameter extraction from CMB: a forecast for
Planck,''
  JCAP {\bf 0803}, 004 (2008)
  [arXiv:0712.2826].

\bibitem{Mangano:2010ei}
  G.~Mangano {\it et al.}, 
  ``Constraining the cosmic radiation density due to lepton number with Big Bang Nucleosynthesis,''
   [arXiv:1011.0916].  
   
   
\bibitem{Fischler:2010xz}
  W.~Fischler and J.~Meyers,
 ``Dark Radiation Emerging After Big Bang Nucleosynthesis?,''
  Phys.\ Rev.\  D {\bf 83}, 063520 (2011)
  [arXiv:1011.3501].
  
\end{thebibliography}
\end{document}